\documentstyle[12pt]{article}
\newcommand{\nc}{\newcommand}
\nc{\be}{\begin{equation}}
\nc{\ee}{\end{equation}}
\nc{\bea}{\begin{eqnarray}}
\nc{\eea}{\end{eqnarray}}

\nc{\eqn}[1]{{(\ref{#1})}}
\nc{\cA}{{\cal A}}
\nc{\cB}{{\cal B}}
\nc{\cC}{{\cal C}}
\nc{\cD}{{\cal D}}
\nc{\cE}{{\cal E}}
\nc{\cF}{{\cal F}}
\nc{\cG}{{\cal G}}
\nc{\cH}{{\cal H}}
\nc{\cI}{{\cal I}}
\nc{\cJ}{{\cal J}}
\nc{\cK}{{\cal K}}
\nc{\cL}{{\cal L}}
\nc{\cM}{{\cal M}}
\nc{\cN}{{\cal N}}
\nc{\cO}{{\cal O}}
\nc{\cP}{{\cal P}}
\nc{\cQ}{{\cal Q}}
\nc{\cR}{{\cal R}}
\nc{\cS}{{\cal S}}
\nc{\cT}{{\cal T}}
\nc{\cU}{{\cal U}}
\nc{\cV}{{\cal V}}
\nc{\cW}{{\cal W}}
\nc{\cX}{{\cal X}}
\nc{\cY}{{\cal Y}}
\nc{\cZ}{{\cal Z}}
\nc{\simo}[1]{{\stackrel{#1}{\simeq}}}
\nc{\geqo}[1]{{\stackrel{#1}{\geq}}}
\nc{\geo}[1]{{\stackrel{#1}{>}}}
\nc{\guo}[1]{{\stackrel{#1}{\succ}}}

\nc{\rbo}{\raisebox}
\nc{\RR} {\rangle \! \rangle}
\nc{\LL} {\langle \! \langle}
\nc{\rmi}[1]{{\mbox{\small #1}}}
\nc{\eq}{eq.~}
\nc{\nr}[1]{(\ref{#1})}
\nc{\ul}{\underline}
\nc{\mc}{\multicolumn}
\nc{\todo}[1]{\par\noindent{\bf $\rightarrow$ #1}}

\title{\begin{flushright} {\small HD-THEP-95-30} \end{flushright}
\vskip 1cm
High temperature critical
O(N) field models by LCE series}
\author{Thomas Reisz\thanks{Email address reisz@ThPhys.Uni-Heidelberg.DE}
        \\Institut
        f\"ur Theoretische Physik,\\
	Universit\"at Heidelberg, \\
        Philosophenweg 16, \\
	D-69120 Heidelberg, Germany}

\begin{document}
\maketitle

\begin{abstract}
The critical properties of renormalizable O(N) field models
are determined by means of the high order ($\geq 18$) behaviour
of convergent linked cluster series on
finite temperature lattices.
It is shown that those models become weakly coupled at the phase
transition.
The critical exponents agree to those of the corresponding
superrenormalizable 3-dimensional models.
Concerning critical amplitudes and
subcritical behaviour, corrections induced
by renormalizable couplings are measurable.
\end{abstract}



\newpage


Quantum field theories at finite temperature are the target of recent
intensive studies. This is not at least because they are supposed to allow
for a quantitative investigation of the physics of
the early universe, including early phase transitions,
as well as of heavy ion collisions.
There is increasing effort to develop and apply techniques and
methods suitable for a serious understanding
both of the theories themselves and of their direct physical
(measurable) implications.
For instance,
standard weak coupling perturbative expansions suffer from
various problems related to severe infrared singularities for theories
with zero mass excitations.
Various proposals exist to resolve or to circumvent those problems.
Among those are hard thermal loop resummations \cite{braaten},
finite volume scaled perturbation theory \cite{red,bengt},
and the techniques of scale dependent flow equations
\cite{christof1,christof2} and block spin renormalization
group \cite{kerres}.

A very promising way is to start with well defined
lattice field theories.
This in turn implies computer aided studies both numerically and
analytically. Convergent expansion methods allow for arbitrary
precise measurements within their convergence domain, limited
only by available skill and resources.
One of those convergent techniques is the (vertex renormalized)
linked cluster expansion (LCE) (e.~g.~\cite{Wortis,ID,LW1}).
It is a series expansion of
correlation functions with respect to the strength fields at different
lattice sites are coupled. The weaker those couplings, the higher
the disorder of the system.
Those techniques have recently been generalized to apply properly to
field theories on finite temperature lattices \cite{thomas}.

In this letter we present results obtained by applying the LCE to
the high temperature phase of (low energy effective)
O(N) symmetric scalar models on the
lattice with renormalizable quartic interaction.
The symmetry restoration for the O(4) case has been investigated
earlier in \cite{kjansen}.
Those models provide one of the fundamental building blocks of
elektroweak standard
models and serve as effective theories for the QCD chiral phase transition.
The expansion parameter is the isotropic nearest neighbour coupling.
The coefficients of susceptibility series depend on the
quartic coupling in a non-trival way. They
are uniform in sign by
inspection, which implies convergence up to the critical
temperature. Decoding the high order behaviour, detailed information
on critical properties is obtained, such as numbers or bounds on universal
critical exponents and amplitudes.

The standard canonical form of the action reads
\be \label{canaction}
   S(\Phi,\kappa,\lambda) =  \sum_{x\in\Lambda} \left(
    \Phi^2 + \lambda (\Phi^2-1)^2 \right) -
   2 \kappa \; \sum_{<xy>} \Phi(x) \cdot \Phi(y),
\ee
where the last sum is over nearest neighbour lattice sites on the
hypercube
$\Lambda = L_0 \times \infty^3$ with periodic boundary conditions
imposed, and $0\leq\lambda\leq\infty$.
We are interested in physical (renormalized) coupling constants
as defined by the large scale behaviour
of the vertex correlations
\be \label{mass}
  \widetilde\Gamma_{ab}^{(2)}(p,-p) = - \frac{1}{Z_R} \,
   ( m_R^2 + p^2 + O(p^4) ) \; \delta_{a,b} \quad
   {\rm as} \;p=(p_0=0,\vec{p}\to 0) ,
\ee
and
\bea
    \label{gren}
  \widetilde\Gamma_{abcd}^{(4)}(p_1=\cdots =p_4 =0) & = &
   - \frac{1}{Z_R^2} \, \frac{g_R}{3} C_4(a,b,c,d), \\
   \label{hren}
  \widetilde\Gamma_{a_1\cdots a_6}^{(6)}(p_1=\cdots =p_6=0) & = &
   - \frac{1}{Z_R^3} \, \frac{h_R}{15} C_6(a_1,\ldots,a_6).
\eea
The $C_n$ denote the O(N)-invariant tensors,
$C_2(a,b) = \delta_{a,b}$, and
\be
  C_{2n}(a_1,\ldots , a_{2n}) = \sum_{i=2}^{2n} \delta_{a_1,a_i} \,
   C_{2n-2}(a_2,\ldots,\widehat a_i,\ldots,a_{2n}),\quad n\geq 2,
   \nonumber
\ee
where ($\;\widehat{ }\;$) implies omission.
In terms of the standard (connected) susceptibilities,
\bea \label{chi}
  \frac{ C_{2n}(a_1,\dots,a_{2n}) }{ (2n-1)!! } \;
   \chi_{2n} & = & \sum_{x_2,\dots ,x_{2n}}
   < \Phi_{a_1}(0) \Phi_{a_2}(x_2) \cdots
      \Phi_{a_{2n}}(x_{2n}) >^{\rm conn}, \\
      \label{mu}
  \mu_2 & = & \sum_x \left( \sum_{i=1}^3 x_i^2 \right)
   < \Phi_1(0) \Phi_1(x) >^{\rm conn}
\eea
(spatial weight), they are given by
\bea \label{vertconn}
  m_R^2 & = & 6 \frac{\chi_2}{\mu_2} = \frac{ Z_R }{ \chi_2 }, \\
  g_R & = & - \left( \frac{6}{\mu_2} \right)^2 \; \chi_4 , \\
  h_R & = & - \left( \frac{6}{\mu_2} \right)^3 \;
   \left( \chi_6 - 10 \frac{\chi_4^2}{\chi_2} \right).
\eea
It is the set of correlations such as \eqn{chi},\eqn{mu} that
LCEs are applied to in the first place. We do not give
technical or graph theoretical details here. For those the
interested reader is referred to \cite{thomas}.
One major point to be noticed that goes far beyond a purely
graph theoretical rearrangement is the introduction of
"1-particle irreducible" susceptibilities\footnote{They are to be
distinguished from the vertex functions leading to
\eqn{mass}-\eqn{hren}, cf. eg. \cite{LW1,thomas}.}.
It is not just that the series representation of the latter are
constructed explicitly. Their explicit use allows for a more precise
measurement of renormalized coupling constants.
This is particularly useful for the computation of couplings
that are related to $\chi_n$ with $n>2$, such as are
$g_R$ and $h_R$ in the present case.
They normally suffer from the problem that their
critical exponents are differences of those of the higher $\chi_n$
of about the same magnitude. This
adds to the equally severe problem that at fixed order increasing $n$
implies decreasing accuracy due to potentiated singularities.
Fortunately, having the (LCE) 1PI susceptibility series at hand,
we can avoid this problem
by defining appropriate ratios and expressing the latter
in terms of the 1PI series directly.

To this end, we first notice that
connected and 1PI susceptibilities are related by
\bea \label{conn1PI}
  \chi_2 & = & \frac{\chi_2^{\rm 1PI}}{1-16\kappa\chi_2^{\rm 1PI}} , \\
   \label{lce.28.11}
  \mu_2 & = & \frac{\mu_2^{\rm 1PI,mod}}
     {(1-16\kappa\chi_2^{\rm 1PI})^2} , \\
  \chi_4 & = & \frac{\chi_4^{\rm 1PI}}
     {(1-16\kappa\chi_2^{\rm 1PI})^4} , \\ \label{lce.29}
  \chi_6 & = &  \frac{1}{(1-16\kappa\chi_2^{\rm 1PI})^6}\;
    \chi_6^{\rm 1PI,mod} ,
\eea
where
\bea
   \mu_2^{\rm 1PI,mod} & = &
    \mu_2^{\rm 1PI}+12\kappa(\chi_2^{\rm 1PI})^2 , \\
   \chi_6^{\rm 1PI,mod} & = &
    \chi_6^{\rm 1PI} + 120\kappa\;
     \frac{(\chi_4^{\rm 1PI})^2}{ 1-16\kappa\chi_2^{\rm 1PI}}.
\eea
In turn, we define the following intermediate functions.
\bea \label{critfunc1}
  \chi_2^{\rm red} \equiv \frac{1}{d_2} & = & \frac{\chi_2}{\chi_2^{\rm 1PI}}
    = \frac{1}{1-16\kappa\chi_2^{\rm 1PI}}
   \simeq \cA_2^{\rm red}
   \left( 1 - \frac{\kappa}{\kappa_c} \right)^{-\gamma}, \\
   \label{critfunc2}
  \mu_2^{\rm red} & = & \frac{\mu_2 d_2}{\chi_2^{\rm 1PI}}
   \simeq \cA_\mu^{\rm red}
   \left( 1 - \frac{\kappa}{\kappa_c} \right)^{-2\nu} , \\
   \label{critfunc3}
  \chi_4^{\rm red} & = &
   \frac{\chi_4^{\rm 1PI}}{(\mu_2^{\rm 1PI,mod})^2 d_2}
   \simeq \cA_4^{\rm red}
   \left( 1 - \frac{\kappa}{\kappa_c} \right)^{-(\gamma+\omega_g)} , \\
   \label{critfunc4}
  \chi_6^{\rm red} & = &
   \frac{\chi_6^{\rm 1PI,mod}}{(\mu_2^{\rm 1PI,mod})^3 d_2}
   \simeq \cA_6^{\rm red}
   \left( 1 - \frac{\kappa}{\kappa_c} \right)^{-(\gamma+\omega_h)} .
\eea
We have indicated the expected leading singular behaviour.
Possible multiplicative
logarithmic corrections are included in the amplitudes $\cA^{\rm red}$.
{}From this we infer that
\bea \label{physcrit}
  m_R^2 & = & \frac{6}{\mu_2^{\rm red}} \sim
   \left( 1 - \frac{\kappa}{\kappa_c} \right)^{2\nu} , \\
  Z_R & = & \frac{ 6 \chi_2^{\rm 1PI}}{\mu_2^{\rm red} d_2} \sim
   \left( 1 - \frac{\kappa}{\kappa_c} \right)^{\nu\eta}, \\
  g_R & = & - 36 \chi_4^{\rm red} d_2 \sim
   \left( 1 - \frac{\kappa}{\kappa_c} \right)^{-\omega_g} , \\
  h_R & = & - 216 \chi_6^{\rm red} d_2 \sim
   \left( 1 - \frac{\kappa}{\kappa_c} \right)^{-\omega_h}.
\eea
Introducing the lattice spacing $a(\kappa,\lambda)$ such that
$T=1/(L_0 a)$ and $T\to T_c$ as
$a(\kappa,\lambda)\to a(\kappa_c,\lambda)$ for fixed $L_0$,
we get in terms of dimensionfull (physical) quantities
\bea \label{tot4}
 \frac{g_{\rm ph}T}{m_{\rm ph}} & \simeq &
  \cA_4 \left( T - T_c \right)^{-\omega_g-\nu} , \\ \label{tot6}
  h_{\rm ph} T^2 & \simeq &
  \cA_6 \left( T - T_c \right)^{-\omega_h} .
\eea

The results presented below are obtained from the 2- and 4-point
suscecptibility series to 18th order (and the 6-point correlation
to 16th order) available for the complete range of quartic couplings.
We expect the accuracy of an $n$th order computations on the
$L_0\times\infty^3$ lattice to be at least comparable
to that one of the order $n-L_0$ on the symmetric $\infty^4$ lattice,
if in both cases the same methods are used.
So at least lattices with $L_0=4,6$ are covered by our
computation.

The determination of the critical couplings has been done along the
following lines (cf.~also \cite{gaunt,guttmann1,guttmann2}).
The above functions are obtained as series representations in the
hopping parameter $2\kappa$, with coefficients depending
nontrivially on the bare coupling constants $\lambda$.
Let us denote by $f_L$ the truncation of the series representation
of
\be
   f(\lambda ,\kappa) = \sum_{m_f\leq\nu} c_\nu(\lambda) (2\kappa)^\nu
\ee
at order $L\geq 0$,
\be
   f_L(\lambda ,\kappa) = \sum_{m_f\leq\nu\leq L}
   c_\nu(\lambda) (2\kappa)^\nu.
\ee
The susceptibilitiy series are convergent within a circle of
radius $2\kappa_c(\lambda)$. By inspection, the coefficients are
of uniform sign, which implies the (strongest) singularity closest to the
origin lies on the positive real axis.
Using this, the leading critical behaviour of $f$,
assumed to be given by
\be \label{critapprox}
   f \simeq \cA \left( 1-\frac{\kappa}{\kappa_c} \right)^{-\omega}
  \quad\mbox{as $\kappa\to\kappa_c-$} ,
\ee
is determined as follows.
The critical coupling $\kappa_c(\lambda)$ where the 2-point susceptibility
diverges is computed as the smallest positive root of $d_2$, using that
$\chi_2^{\rm 1PI,c}(\lambda)$ $=$
$\lim_{\kappa\to\kappa_c(\lambda)-} \chi_2^{\rm 1PI}(\lambda,\kappa)$
stays finite at $\kappa_c$.
That is,
\bea \label{critkappa}
  & & d_{2,L}( \lambda , \kappa_c(\lambda,L) ) = 0 , \\
  & & \kappa_c(\lambda ,L) = \kappa_c(\lambda)
   + \frac{\delta}{L} + o(L^{-1}) .
\eea
For the regression, we always choose $L\geq 9$.
The values obtained agree within the error bars to those ones got
by the standard ratio criterions on the series representation of
$\chi_2$. In this way, however, the error bars themselves
are considerably smaller.
A possible antiferromagnetic singularity at $-\kappa_c(\lambda)$
is shifted to $-\infty$ by changing to the new variables
$z = \frac{2\kappa}{1-\frac{\kappa}{\kappa_c(\lambda)}}$,
leading to
\be
   \overline{f}_L(\lambda ,z) = \sum_{\nu\leq L}
   \overline{c}_\nu(\lambda) z^\nu.
\ee
Knowing the critical point $\kappa_c$ or $z_c$,
the critical exponent $\omega$ is
obtained by the large order $\nu$ behaviour of the ratio
\be \label{ratio}
  r_\nu = \frac{\overline{c}_\nu}{\overline{c}_{\nu-1}}
  = \frac{1}{z_c} \left( 1 + \frac{\omega-1}{\nu} + R_\nu \right) ,
\ee
where $R_\nu=o(\nu^{-1})$ as $\nu\to\infty$.
The decay strength of $R_\nu$ is sensitive to the presence
of multiplicative logarithmic corrections. A constant
amplitude $\cA$ as above, \eqn{critapprox}, implies that
the large order behaviour of $R_\nu$ is
determined by the subleading power like behaviour on the
boundary of the convergence domain, $R_\nu=O(\nu^{-1-\Delta})$,
where typically $1/2 \leq \Delta \leq 1$
\cite{guttmann1,wegener}.
On the other hand, if $\cA$ diverges or approaches $0$ logarithmically
as $\kappa$ approaches $\kappa_c$,
\be
  \cA \sim \left( \log{ \vert 1-\frac{\kappa}{\kappa_c} \vert }
   \right)^\tau,
\ee
with $\tau\not=0$,
the decay is much weaker,
$R_\nu=O(\frac{1}{\nu\psi(\nu)})=O(\frac{1}{\nu \ln{\nu}})$.
Direct measurement of such logarithmic exponents can hardly be done from
the LCE series. But the $\chi^2/df$ can strongly hints this property.

The critical amplitude is obtained by comparing the series of $f$ to
the right hand side of \eqn{critapprox} and linear regression
$\cA_\nu = \cA + O(\nu^{-1})$.

We don't use Pade' approximants. They give hints to the values of
critical exponents,
but the error bars never happened to be smaller.

Tab.~\ref{o1} and \ref{o4} show the resulting critical line and the critical
exponent $\gamma$ both for the 1-component and the O(4) model,
as obtained on the $4\times\infty^3$ lattice.
The coupling $\overline\lambda$ as defined by
$\overline\lambda = -\frac{N+2}{6}\;\frac{v_4-3(v_2)^2}{(v_2)^2}$
for O(N), with
\be
   v_{2n} = \frac{ \int d^N\Phi \Phi_1^n
    e^{-(\Phi^2 + \lambda (\Phi^2-1)^2)} }
      { \int d^N\Phi
         e^{-(\Phi^2 + \lambda (\Phi^2-1)^2)} },
\ee
is monotonically increasing with $\lambda$ and
$0\leq\overline\lambda\leq 1$ as $0\leq\lambda\leq\infty$.
Beyond the gaussian (zero temperature) fixed point at $\overline{\lambda}=0$
there is definite non-mean field behaviour at larger couplings.
We identify the finite temperature phase transition as the strong
coupling part, including a range
$\overline{\lambda}_{\rm min}$ $\leq$
$\overline{\lambda}$ $\leq 1$
where the critical exponents are stable.
Tab.~\ref{critexp} and \ref{critamp} summarize the critical data
on the exponents and amplitudes for $N=1,\ldots ,4$, obtained
from the high $T$ part of the phase diagram.
They are compared to the data of the corresponding
superrenormalizable 3-dimensional models with the numbers
that we have determined
along the same lines as outlined above\footnote{Cf. also
e.g.~\cite{kanaya,baker}.
For an extended list
of references on 3-dimensional critical numbers both for O(N) field
models and N-vector spin models cf.~\cite{christof2}
and \cite{butera}. For the latter,
the best of those data are given in \cite{butera}.}.

The results can be summarized as follows.
The renormalized quartic coupling vanishes at $T_c$ proportional
to the mass ($\omega_g=-\nu$).
Within bounds, the 6-point coupling has a vanishing critical exponent
($\omega_h=0$).
The most precise data for the critical exponents
$\gamma$ and $\nu$ are obtained by \eqn{ratio} with
$R_\nu=O(\frac{1}{\nu \ln{\nu}})$, indicating multiplicative logarithmic
corrections to \eqn{critapprox}.
All the critical exponents, i.e.~$\gamma$, $\nu$ as well as
$\omega_g$, $\omega_h$ agree in value to those of the 3-dimensional
O(N) $\Phi^4$ models.
The numbers on the critical couplings and exponents
are the most precise ones as yet obtained.
They are in reasonable agreement to earlier investigations
both by Monte Carlo methods \cite{kjansen} and
scale dependent flow equations \cite{christof2}.

For $\omega_g=-\nu$ and $\omega_h=0$, the ansatz \eqn{critapprox}
yields values on the critical amplitudes \eqn{tot4},\eqn{tot6}
summarized in Tab.~\ref{critamp}.
Although the error bars are rather large on those
3rd generation quantities (after the critical line and the
critical exponents), we measure deviations of the 3d data.
This is in conformity with the observation that multiplicative logarithmic
corrections are present that are not provided by the 3d superrenormalizable
models.
Hence we have a twofold strong evidence that a quantitative description
beyond the leading critical behaviour by a 3d effective theory needs
non-superrenormalizable couplings such as $\Phi^6$,
regardless that the quartic coupling vanishes at $T_c$.
This might become important outside of the critical region, and
in particular for first order transitions.

All the 4d data presented above are obtained on the $4\times\infty^3$
lattice.
Consistent results are obtained on the larger $6\times\infty^3$
hypercube.
Yet the relative error bars become about twice as large due to a reduction
of the effective order by two.
As yet, the data for temporal extent $L_0\geq 8$
are not very predictive to 18th order of computations.
There are future ways to cure this.
First, the construction algorithms are that efficient that
the order of the susceptibility series can still be increased
within available resources. Second, zero temperature
"background" can be subtracted. Third, convergence behaviour can
be improved by including "irrelevant" next to nearest neighbour
interactions.

%
%




\begin{table}[htb]
\caption{\label{o1} The critical line $\kappa_c(\lambda)$
for the 1-component $\Phi^4$
model on the $4\times\infty^3$ lattice.
Also shown is the critical exponent $\gamma(\lambda)$, that exhibits
the high temperature phase transition as the strong coupling range
of the phase diagram.
}
\vspace{0.5cm}

\begin{center}

\begin{tabular}{|l|l|l|l|}
\hline \hline
$\;\overline{\lambda}$ & $\quad\lambda$ & $\qquad\kappa_c$ & $\qquad\gamma$ \\
[0.5ex] \hline

 1   &$\infty$&0.075773(21) &1.2407(83)  \\
 0.95&6.92952 &0.085272(24) &1.2392(87)   \\
 0.9 &4.3303  &0.093339(27) &1.2406(102)  \\
 0.85&3.25072 &0.10085(4)   &1.2395(106)  \\ [0.5ex] \hline
 0.8 &2.5836  &0.10789(4)   &1.2372(97)  \\
 0.7 &1.7320  &0.12053(4)   &1.2387(110)  \\
 0.6 &1.1769  &0.13094(4)   &1.1977(85)   \\
 0.5 &0.77841 &0.13845(3)   &1.1731(67)   \\
 0.4 &0.48548 &0.14244(3)   &1.1570(58)   \\
 0.3 &0.27538 &0.14239(3)   &1.1323(65)   \\
 0.2 &0.13418 &0.13860(2)   &1.1007(44)   \\
 0.1 &0.04770 &0.13225(2)   &1.0538(46)   \\
 0.05&0.01997 &0.12866(1)   &1.0311(24)   \\
 0.01&0.003457&0.12573(1)   &1.0066(24)   \\ \hline
\end{tabular}

\end{center}

\end{table}


\begin{table}[htb]
\caption{\label{o4} The same as Tab.~1, but for the
O(4) symmetric model.
}
\vspace{0.5cm}

\begin{center}

\begin{tabular}{|l|l|l|l|}
\hline \hline
$\;\overline{\lambda}$ & $\quad\lambda$ & $\qquad\kappa_c$ & $\qquad\gamma$ \\
[0.5ex] \hline

 1    & $\infty$ & 0.30967(16) & 1.4473(202)  \\
 0.975&39.4673   & 0.30800(18) & 1.4516(229)  \\
 0.95 & 19.4314  & 0.30614(19) & 1.4417(243)  \\ [0.5ex] \hline
 0.925& 12.725   & 0.30418(20) & 1.4310(258)  \\
 0.9  & 9.34653  & 0.30211(18) & 1.4211(233)  \\
 0.8  & 4.12700  & 0.29204(17) & 1.3780(236)  \\
 0.7  & 2.24446  & 0.27772(15) & 1.3380(206)  \\
 0.6  & 1.25586  & 0.25802(13) & 1.2982(197)  \\
 0.5  & 0.67905  & 0.23313(11) & 1.2569(184)  \\
 0.4  & 0.34514  & 0.20532(8)  & 1.2164(153)  \\
 0.3  & 0.16348  & 0.17848(6)  & 1.1728(123)  \\
 0.2  & 0.07041  & 0.15583(4)  & 1.1247(101)  \\
 0.1  & 0.02370  & 0.13826(2)  & 1.0696(56)   \\
 0.05 & 0.00989  & 0.13116(1)  & 1.0346(25)   \\ \hline
\end{tabular}

\end{center}

\end{table}


\begin{table}[htb]
\caption{\label{critexp} The critical exponents of the $\Phi^4$
O(N) models for $N=1,\ldots ,4$, both in 4 dimensions
(upper values) and in 3 dimensions at zero temperature
(lower values).
}
\vspace{0.5cm}

\begin{center}

\begin{tabular}{|c|c|l|l|l|l|l|}
\hline \hline
N & $\overline{\lambda}_{\rm min}$ & $\qquad\gamma$ & $\qquad\nu$ &
 $\qquad\nu\eta$ & $\qquad\omega_g$ & $\qquad\omega_h$ \\
[0.5ex] \hline

   & 0.85 & 1.2400(87) & 0.6300(49) & 0.0193(132)& -0.6296(95)
& -0.0016(136) \\ \cline{2-2}
 \raisebox{1.5ex}[-1.5ex]{1}
   & 0.90 & 1.2406(36) & 0.6301(18) & 0.0183(52) & -0.6300(42)
& -0.0008(72) \\ \hline
   & 0.90 & 1.3238(139) & 0.6694(64)& 0.0206(150)& -0.6683(131)
& $\;$0.0030(191) \\ \cline{2-2}
 \raisebox{1.5ex}[-1.5ex]{2}
   & 0.90 & 1.3250(52) & 0.6734(28) & 0.0199(76) & -0.6788(75)
& $\;$0.0040(126) \\ \hline
   & 0.90 & 1.4032(156) & 0.7167(76)& 0.0246(169)& -0.7132(164)
& $\;$0.0010(262) \\ \cline{2-2}
 \raisebox{1.5ex}[-1.5ex]{3}
   & 0.90 & 1.4029(85) & 0.7131(40) & 0.0232(87) & -0.7165(115)
& -0.0167(218) \\ \hline
   & 0.95 & 1.4469(225)& 0.7356(93) & 0.0257(200)& -0.7285(276)
& $\;$0.0075(397) \\ \cline{2-2}
 \raisebox{1.5ex}[-1.5ex]{4}
   & 0.90 & 1.4504(113)& 0.7361(68) & 0.0213(123) & -0.7268(134)
&-0.0104(247) \\ \hline
\end{tabular}

\end{center}

\end{table}


\begin{table}[htb]
\caption{\label{critamp} The corresponding critical amplitudes
$\cA_4$ and $\cA_6$ for
$\omega_g=-\nu$ and $\omega_h=0$, both in 4 dimensions (upper values)
and in 3 dimensions (lower values).
The numbers hold under the assumption that no multiplicative logarithmic
factors occur.
}
\vspace{0.5cm}

\begin{center}

\begin{tabular}{|c|r|r|}
\hline \hline
N & $\cA_4\qquad$ & $\cA_6\quad$ \\
[0.5ex] \hline

   & 16.61 $\pm$ 1.46 & 590 $\pm$ 103 \\
 \raisebox{1.5ex}[-1.5ex]{1}
   & 23.72 $\pm$ 1.49 & 1386 $\pm$ 174 \\ \hline
   & 15.39 $\pm$ 2.24 & 464  $\pm$ 121 \\
 \raisebox{1.5ex}[-1.5ex]{2}
   & 20.33 $\pm$ 1.76 & 913 $\pm$ 177 \\ \hline
   & 14.99 $\pm$ 2.40 & 406 $\pm$ 114 \\
 \raisebox{1.5ex}[-1.5ex]{3}
   & 17.87 $\pm$ 1.68 & 672 $\pm$ 142 \\ \hline
   & 13.83 $\pm$ 2.37 & 293 $\pm$ $\;\;$94 \\
 \raisebox{1.5ex}[-1.5ex]{4}
   & 15.29 $\pm$ 1.56 & 447 $\pm$ 105 \\ \hline
\end{tabular}

\end{center}

\end{table}

\end{document}